\documentclass[aps,prd,reprint,twocolumn,nofootinbib,superscriptaddress,showpacs,floatfix,preprintnumbers]{revtex4-1}

\usepackage[dvipsnames]{xcolor}     
\definecolor{lcolor}{rgb}{0.5,0,0}
\definecolor{citcolor}{rgb}{0,0,1}
\usepackage[breaklinks,colorlinks,urlcolor=blue,citecolor=citcolor,linkcolor=lcolor,linktoc=all]{hyperref}
\usepackage{color}
\usepackage{graphicx}	
\usepackage[utf8]{inputenc}
\usepackage{amsmath} 
\usepackage{amssymb}
\usepackage[dvipsnames]{xcolor}      

\newcommand{\be}{\begin{equation}}
\newcommand{\ee}{\end{equation}}
\newcommand{\la}{\lambda}

\begin{document}

\preprint{APCTP Pre2023 - 007}

\title{Holographic baryons, dense matter and neutron star mergers}

\author{Matti J{\"a}rvinen}
\affiliation{Asia Pacific Center for Theoretical Physics,  Pohang 37673, Republic of Korea}  
\affiliation{Department of Physics, Pohang University of Science and Technology Pohang 37673, Republic of Korea}

\begin{abstract}
The gauge/gravity duality, combined with information from lattice QCD, nuclear theory, and perturbative QCD, can be used to constrain the equation of state of hot and dense QCD. I discuss an approach based on the holographic V-QCD model. I start by reviewing the results from the construction of the V-QCD baryon as a soliton of the gauge fields in the model.
Then I discuss implementing nuclear matter in the model by using a homogeneous approach. The model predicts a strongly first order phase transition from nuclear to quark matter with a critical endpoint. By using the model in state-of-the-art simulations of neutron star binaries with parameters consistent with GW170817, I study the formation of quark matter during the merger process. 
\end{abstract}

\maketitle

\section{Introduction and motivation}

Exploring QCD at finite density and temperature is challenging both theoretically and experimentally~\cite{Brambilla:2014jmp}.  
Producing hot QCD plasma in the laboratory is complicated due to the high characteristic temperature of QCD, $1~\mathrm{GeV} \approx 10^{13}$~K, but can be achieved in high energy heavy-ion collisions. However the density of the produced plasma is typically low compared to the density of nuclear matter. There are ongoing efforts to obtain results also at higher densities: the beam energy scan at RHIC is already probing the region where the critical end point of the nuclear to quark matter transition is expected to lie, and future experiments such as FAIR and NICA will push the results to higher densities, even above the nuclear saturation density of $n_s \approx 0.15$~fm$^{-3}$.

\begin{figure} 
 \centerline{\includegraphics[width=0.45\textwidth]{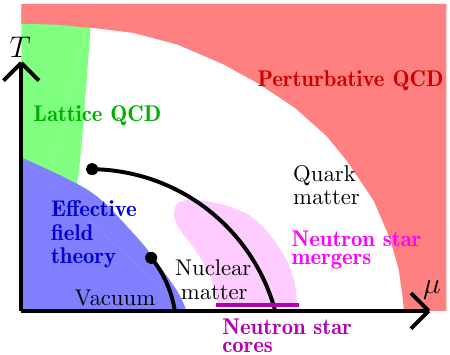}}
 \caption[]{A sketch of the QCD phase diagram at finite temperature and density. The green, blue, and red  regions show roughly where lattice analysis, various effective field theory methods, and perturbative QCD can be trusted, respectively. Figure adapted from~\protect\cite{Jarvinen:2021jbd}.}
 \label{fig1}
\end{figure}

Theoretical analysis of hot and dense QCD is likewise challenging and the used approaches have limitations.
Various first-principles methods cover different parts of the phase diagram of QCD at finite temperature and density as depicted in Fig.~\ref{fig1}. 
\begin{itemize}
                                                                                                                                                 \item Lattice methods~\cite{Ding:2015ona} (green region) only work at low densities due to the sign problem~\cite{deForcrand:2010ys}. 
                                                                                                                                                 \item Perturbative QCD analysis (red region) is reliable at asymptotically high densities and temperatures~\cite{Kurkela:2014vha,Gorda:2021znl}. 
                                                                                                                                                 \item At low densities and temperatures, the relevant degrees of freedom in QCD are the hadrons, which can be described in terms of various effective theory methods, such as chiral perturbation theory and mean field theory approximations (see, e.g.,~\cite{Drischler:2017wtt,Tews:2018iwm,Keller:2020qhx}). 
                                                                                                                                                \end{itemize}
In the white region of intermediate densities, however, no reliable first-principles methods are available. This region  is of considerable interest: It includes the conjectured phase transition from hadronic to quark matter phase. Also, matter in the centers of neutron stars as well as in the dense regions of neutron star mergers is known to lie in this region. Therefore theoretical progress at these intermediate densities is urgently needed.

The fact that neutron star densities lie in the white region of Fig.~\ref{fig1} also means that observations of neutron stars and neutron star mergers can be used to constrain QCD in this region (see, e.g.~\cite{Oertel:2016bki}). Neutron stars, from the QCD point of view, are blobs of cold and dense QCD matter at rest: to first approximation effects of finite temperature, rotation and magnetic fields can be neglected. The structure of an isolated star is determined by the Tolman-Oppenheimer-Volkoff (TOV) equations, which map the equation of state (EOS) of QCD to the mass-radius relation of neutron stars. Therefore, measurements of masses and radii of neutron stars can in principle be converted to measurements of the EOS. Current measurements of neutron star masses have however large uncertainties, and the results for the radii are even less certain. Due to these uncertainties, the measurements currently only give limited information on the EOS. The main constraint is coming from the  Shapiro delay measurements of the most massive known stars (e.g.  J0348+0432 and J0740+6620~\cite{Antoniadis:2013pzd,Cromartie:2019kug,Fonseca:2021wxt}): the EOS must be such that it supports masses of at least twice the solar mass, $M_\mathrm{max} \gtrsim 2 M_\odot$. But in future, improved estimates of masses and radii of neutron stars (e.g. form the NICER experiment) are expected to give significantly stronger constraints for the EOS.

Apart from measurements of isolated neutron stars, the observations of neutron star mergers give complementary information on the structure of neutron stars. Currently, by far the best available data is from the ``multi-messenger'' measurements of the event GW170817~\cite{TheLIGOScientific:2017qsa,GBM:2017lvd}. These measurements give information, among other things, on the tidal deformability parameter $\Lambda$, which measures how strongly neutron star deform in strong gravitational field. The observation of the gravitational waves during the inspiral phase of the merger sets the bound $\Lambda \lesssim 580$ (assuming the same EOSs for both neutron star constituents) at 90\% confidence level~\cite{Abbott:2018exr}. This translates roughly an upper bound  of about 13.5 km for the radius of neutron stars having the mass $\approx 1.4~M_\odot$.

In the absence of reliable theoretical estimates at intermediate densities, one can use model independent interpolations to study how these neutron star measurements constrain the QCD EOS~\cite{Kurkela:2014vha,Annala:2017llu,Most:2018hfd,Komoltsev:2021jzg}. That is, one considers all possible EOSs which agree with theoretical computations from chiral perturbation theory at low densities, perturbative QCD at high densities, are physically reasonable, and agree with the observations within their uncertainties. Such interpolated EOSs have been used to study the effect of the above limits on $M_\mathrm{max}$ and $\Lambda$~\cite{Annala:2017llu,Most:2018hfd} and also the effect of other measurements~\cite{Annala:2019puf,Annala:2021gom,Altiparmak:2022bke}. After taking into account all the constraints, a sizable uncertainty in the EOS still remains, even at zero temperature. Going to finite temperature, in the region relevant for neutron star mergers, uncertainties will grow. 

A new approach which could reduce such uncertainties is  to use gauge/gravity duality as a guideline. Indeed, several attempts in this direction have been carried out recently by using various models, such as D3-D7~\cite{Hoyos:2016zke,Annala:2017tqz,BitaghsirFadafan:2019ofb}, Witten-Sakai-Sugimoto~\cite{Kovensky:2020xif,Pinkanjanarod:2020mgi,Kovensky:2021kzl}, Einstein-Maxwell~\cite{Ghoroku:2019trx,Mamani:2020pks,Ghoroku:2021fos} and hard wall~\cite{Bartolini:2022rkl} models. See also the reviews~\cite{Jarvinen:2021jbd,Hoyos:2021uff}. Here I will focus on results obtained by using the V-QCD model~\cite{Jarvinen:2011qe}.

\section{V-QCD and quark matter}

V-QCD is a bottom-up holographic model for QCD which aims at precise modeling of the properties of QCD both by using inspiration from string theory and by fitting QCD data~\cite{Jarvinen:2011qe} (see the review~\cite{Jarvinen:2021jbd} for more details). In the name of the model, the letter V refers to the Veneziano limit~\cite{Veneziano:1976wm}:
\be
  N_c\ , \ N_f \to \infty \quad \mathrm{with} \quad x \equiv \frac{N_f}{N_c} \quad \mathrm{fixed.}
\ee
When fitting to data however $N_f$ and $N_c$ will be set to their physical values ($2$ or $3$), which helps to capture the $1/N$ corrections to observables such as the EOS.

The model is composed of two building blocks:
\begin{enumerate}
 \item Improved Holographic QCD (IHQCD), a string-inspired model for the Yang-Mills theory, defined through an adjusted five-dimensional Einstein-dilaton gravity~\cite{Gursoy:2007cb,Gursoy:2007er}.
 \item A method for introducing flavors and chiral symmetry breaking via a tachyonic brane setup~\cite{Bigazzi:2005md,Casero:2007ae}.
\end{enumerate}
The dictionary contains various fields, but the most important degrees of freedom are the two scalars:
\begin{itemize}
 \item The dilaton $\lambda = e^{\phi}$, which arises from the IHQCD sector, and is dual to the $\mathrm{tr}F^2$ operator in QCD. The source is therefore the 't Hooft coupling, which explains the notation: near the boundary, the source term dominates and the bulk field $\lambda$ can be identified with the 't Hooft coupling.
 \item The tachyon $\tau$, which arises from the flavor sector, and is dual to the quark bilinear operator $\bar \psi \psi$. The source of field is therefore the quark mass. The condensation of the tachyon in the bulk implies chiral symmetry breaking of the field theory at the boundary.
\end{itemize}
The action, including only terms needed for the computation of the phase diagram and the EOS, is given by
\begin{widetext}
  \begin{align} \label{eq:IHQCD}
 \mathcal{S}_\mathrm{V-QCD} &= \mathcal{S}_\mathrm{IHQCD}+\mathcal{S}_\mathrm{DBI}
 = {N_c^2} M^3  \int d^5x\, \sqrt{-\det g}\Bigg[R-\frac{4}{3}\frac{(\partial \la)^2}{\la^2} +V_g(\la)\Bigg] - \\ 
 & \ \ \  - {N_fN_c} M^3\!\! \int\!\! d^5 x\, {V_{f0}}(\la)e^{-\tau^2}\!\!  \sqrt{\!-\!\det(g_{ab}\!+\! {\kappa}(\la) \partial_a \tau \partial_b \tau \!+\! {w}(\la)  F_{ab})} \label{eq:DBI}
\end{align}
\end{widetext}
where the first term is the action for the glue sector of the model, i.e., closed strings, and matches with the basic version of  action of IHQCD. The second term is the action for the flavor sector: a generalized tachyonic Dirac-Born-Infeld action, motivated by a brane setup with a space filling pair of $D4$ branes. We also included an Abelian gauge-field appearing through the field strength $F_{ab}$. In the Veneziano limit (and also for physical values of $N_f$ and $N_c$) the flavor sector is fully backreacted to the glue, as both actions are $\mathcal{O}(N^2)$. Due to the backreaction, there is not string theory derivation of the DBI action. Therefore we have generalized the action by including three potential functions depending on the dilaton: $V_{f0}$, $\kappa$, and $w$, which need to be fixed to finalize the definition of the model.

The thermodynamics (assuming zero quark mass for simplicity) is determined by using the standard gauge/gravity dictionary, analyzing possible regular homogeneous (in space-time coordinates) saddle-points of the action~\cite{Alho:2012mh,Alho:2013hsa,Alho:2015zua}. Two phases appear, one phase given by a horizonless ``thermal gas'' geometry with condensed tachyon, and the other phase given by a black hole geometry with vanishing tachyon. The thermal gas phase is essentially empty space, and thermodynamics is trivial. The condensation of the tachyon is driven by the exponential term in the DBI action~\eqref{eq:DBI}.  The thermodynamics of the black hole phase is determined through black hole thermodynamics: the entropy density is given by the area of the black hole, and temperature is the surface gravity. Moreover, baryon number chemical potential is turned on as the boundary value of the temporal component of the gauge field: $\mu = \hat A_t|_\mathrm{bdry}$. 

In order to fully determine the model the functions $V_g$, $V_{f0}$, $\kappa$, and $w$ need to be fixed. The asymptotic behavior of these functions at small and large values of the dilaton field is largely determined by qualitative arguments. That is,
requiring that the model has asymptotic freedom, discrete spectrum, linear confinement, reasonable phase diagram at finite  density, qualitatively correct hadron spectrum at finite quark mass, and regular solutions at finite $\theta$-angle fixes these asymptotics up to a few parameters~\cite{Gursoy:2007cb,Gursoy:2007er,Jarvinen:2011qe,Arean:2013tja,Jarvinen:2015ofa,Arean:2016hcs,Ishii:2019gta}. The remaining degrees of freedom are then pinned down by comparing to lattice data for thermodynamics at low densities~\cite{Gursoy:2009jd,Jokela:2018ers} or by comparing to hadron spectra~\cite{Amorim:2021gat} or by doing both simultaneously~\cite{Jarvinen:2022gcc}. This completes the description of quark gluon plasma and quark matter in the model.

\begin{figure*} 
 \centerline{\includegraphics[width=0.65\textwidth]{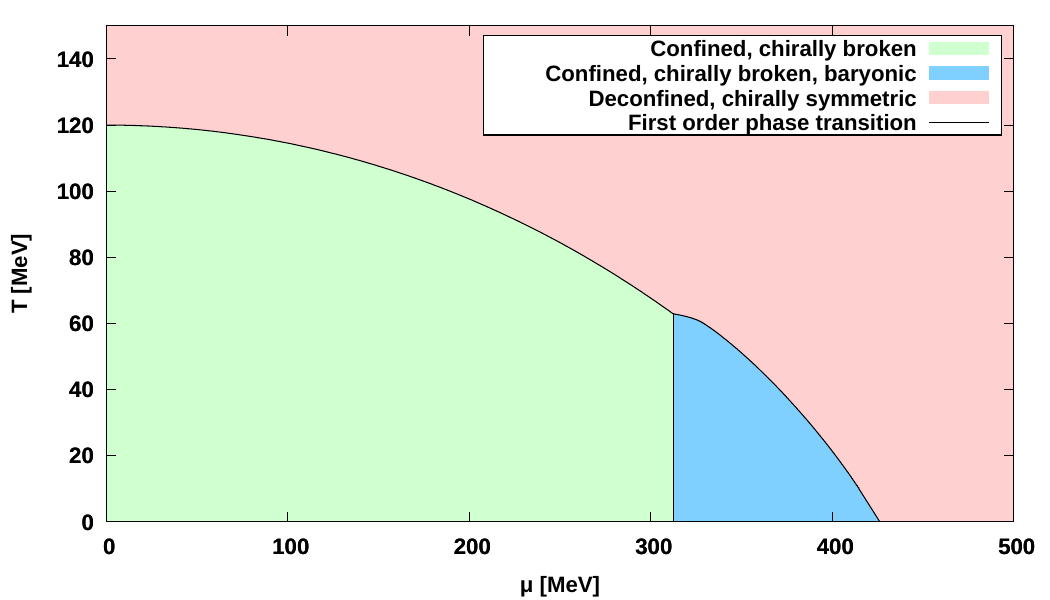}}
 \caption[]{The phase diagram of V-QCD for potentials 7a and including homogeneous nuclear matter. From~\protect\cite{Ishii:2019gta}.}
 \label{fig2}
\end{figure*}

After this analysis, most of the phase structure shown in Fig.~\ref{fig2} can be drawn. Here the green and the red phases are the thermal as and black hole phases, respectively. They are separated by a first order Hawking-Page transition. The blue phase will be discussed in Sec.~\ref{sec:NM}. The precise potentials used for this plot are the set 7a given in~\cite{Jokela:2018ers,Ishii:2019gta}, which were fitted to lattice data for the thermodynamics of pure Yang-Mills at large $N_c$~\cite{Panero:2009tv} and QCD with 2+1 flavors from~\cite{Borsanyi:2011sw,Borsanyi:2013bia}. Interestingly, also the EOS extrapolated to zero $T$ and finite $\mu$ is consistent with constraints from perturbation theory in the regime of intermediate densities~\cite{Jokela:2018ers}.

\section{Baryons in V-QCD}\label{sec:soliton}

The first step towards implementing nuclear matter is to  construct a holographic dual for a single baryon. Baryons are special objects at large $N$ and therefore in holographic models. Indeed, their masses are proportional to $N_c$~\cite{Witten:1979kh}, and in large $N_c$ effective theory they can be described as topological solitons of pion fields, Skyrmions~\cite{Skyrme:1961vq}. In the original AdS/CFT setup proposed by Maldacena, baryons are obtained by wrapping a $D5$ brane around the internal $S^5$ of the AdS$_5 \times S^5$~\cite{Witten:1998xy}. This D-brane acts as a ``baryon vertex'', a source for $N_c$ fundamental strings that represent the quarks. 

In a top-down setup with flavor branes, the strings sourced by the vertex end on these branes. The best studied  example is the Witten-Sakai-Sugimoto (WSS) D4-D8-$\overline{\mathrm{D8}}$ setup, where the baryon vertex is obtained by wrapping a D4 brane around the internal S$^4$ space~\cite{Kim:2006gp,Hata:2007mb,Bolognesi:2013nja}. The strings pull the vertex on the flavor brane, where the dissolved D4 brane is implemented as a soliton of the flavored, non-Abelian gauge fields. Such as soliton can be shown to be dual to a skyrmion-like soliton at the boundary, and becomes similar to the Belavin-Polyakov-Schwartz-Tyupkin (BPST) soliton of Yang-Mills theory~\cite{Belavin:1975fg} in the limit of strong coupling. The soliton is time-independent, and localized in spatial and holographic directions.

In bottom-up models, the natural way to implement baryons is therefore to consider solitons of non-Abelian gauge fields. Indeed, soliton solutions have been found numerically in hard wall models~\cite{Pomarol:2007kr,Pomarol:2008aa,Gorsky:2012eg,Gorsky:2013dda,Gorsky:2015pra}. The solutions in these models however have some shortcomings:
\begin{itemize}
 \item In the WSS model, the size of the soliton is suppressed by a factor $1/\sqrt{\la}$ in the limit of strong coupling, which is required for the supergravity description to be reliable. Also, the interplay of the soliton with the chiral symmetry breaking is difficult to study.
 \item In the hard wall models, the solitons are centered at the IR cutoff (the ``hard wall'') and their properties depend on the IR setting in an ad-hod manner.
\end{itemize}
These issues can be at least mostly fixed in the V-QCD setup. In V-QCD at zero quark mass, there is essentially only one energy scale, as one also expects to be the case in QCD, and this scale will determine the baryon size. The interplay of the baryon with chiral symmetry breaking is explicit due to the coupling of the tachyon field and the soliton. Moreover, the soliton is centered at a location in the bulk which is determined by dynamics rather than set by hand.

An essential ingredient in the description of the baryon is the Chern-Simons (CS) term of the flavor branes. In bottom-up setups the baryon normally contains both left and right handed non-Abelian gauge fields $A_{L/R}$, which are dual to left and right handed currents $\bar \psi \gamma^\mu \tau^a(1\pm \gamma_5)\psi$ and will appear in the CS term. Schematically, the term takes the form
\be
 S_\mathrm{CS} \propto N_c \int dt \hat A_t \wedge \mathrm{tr}\left[F_L\wedge F_L- F_R\wedge F_R\right] \ .
\ee
Therefore, the instanton number density of the soliton, described by the $F \wedge F$ factors, gives rise to a baryon number, i.e., the charge corresponding to the field $\hat A_t$.

We did not include the CS term in~\eqref{eq:IHQCD} and~\eqref{eq:DBI} because it does not contribute to the phase diagram or the EOS when only quark matter is included. For the method we are using to implement the flavors, i.e., the tachyonic space filling branes, it has been solved using an approximation scheme in~\cite{Casero:2007ae} in flat-space boundary string field theory approach, and has quite complicated form. However for a general bottom-up approach it is desirable to derive an expression that is only constrained by symmetry, and contains the full flavored tachyon field $T^{ij}$ (dual to $\bar \psi^i \phi^j$ where $i$, $j$ are the flavor indices). We derived such a general form in~\cite{Jarvinen:2022mys}, assuming that the tachyon has the form $T^{ij} =\tau U^{ij}$ where $\tau$ is real and $U$ is unitary. The result is an integral over an five-form $\Omega_5$, which decomposes into three terms:
\be \label{eq:Omega5sol}
 \Omega_5(\tau,U,A_L,A_R) = \Omega_5^0 + \Omega_5^c +dG_4 \ .
\ee
This expression is constrained among other things by the requirement that its gauge transformation is a boundary term, and matches with the flavor anomalies of QCD in the same way as the flat-space expression found in~\cite{Casero:2007ae}. This means in particular that the differential of the gauge transformation vanishes, $d \delta \Omega_5 =0$, which is the most important condition when solving for $\Omega_5$. This condition is reflected in the properties of the three terms in~\eqref{eq:Omega5sol}:
\begin{itemize}
 \item $\Omega_5^0$ is the most general \emph{gauge-invariant} five form that also has the expected signatures under parity and charge conjugation, so $\delta \Omega_5^0 = 0$. It is given as a sum of four terms which are individually gauge invariant:
 \be
  \Omega_5^0 = \sum_i^{4} f_i(\tau) \Omega_i^0(U,A_L,A_R)
 \ee
 where functions $f_i$ are otherwise arbitrary but their values at $\tau=0$ are know and they should vanish at $\tau \to \infty$.
 \item The ``pure gauge'' term $\Omega_5^c$ is closed but not exact:
 \be 
  \Omega_5^c = \mathrm{Tr} (U^\dagger dU)^5 \ .
 \ee
 \item The boundary term $dG_4$ is complicated but completely fixed by the flavor anomalies.
\end{itemize}
As the result is a sum of closed and gauge-invariant terms, indeed trivially $d \delta \Omega_5 =0$. Interestingly, the closed expression $\Omega_5^c +dG_4$ matches with the gauged effective Wess-Zumino Lagrangian for QCD discussed in~\cite{Witten:1983tw,Kaymakcalan:1983qq,Manes:1984gk}.

After finding the CS term, the setup for the baryon solution is almost complete. It turns out that perhaps surprisingly (at least at zero quark mass) the baryon number is completely fixed by the $dG_4$ boundary term and receives no contributions from the gauge-invariant term. The functions $f_i(\tau)$ however do affect the shape of the soliton, and we choose to use the flat-space expressions~\cite{Casero:2007ae} for them. 

As I pointed out above, the precise definition of the V-QCD model requires determining several potentials, typically by fitting lattice or experimental data. Such fits were carried out in~\cite{Jokela:2018ers}, by mostly using lattice data for thermodynamics, or in~\cite{Amorim:2021gat}, by using a large amount of experimental data for mesons masses, including highly exited states (see also~\cite{Ballon-Bayona:2017vlm,Amorim:2018yod} which did a similar analysis for the Yang-Mills theory). In order to obtain a reasonable background for the baryon solution, we carried out a new fit where the predictions of the model were compared to both lattice thermodynamics and key observables of hadrons, including masses of lowest lying mesons and the pion decay constant~\cite{Jarvinen:2022gcc}. 

\begin{figure*} 
 \centerline{\includegraphics[width=0.49\textwidth]{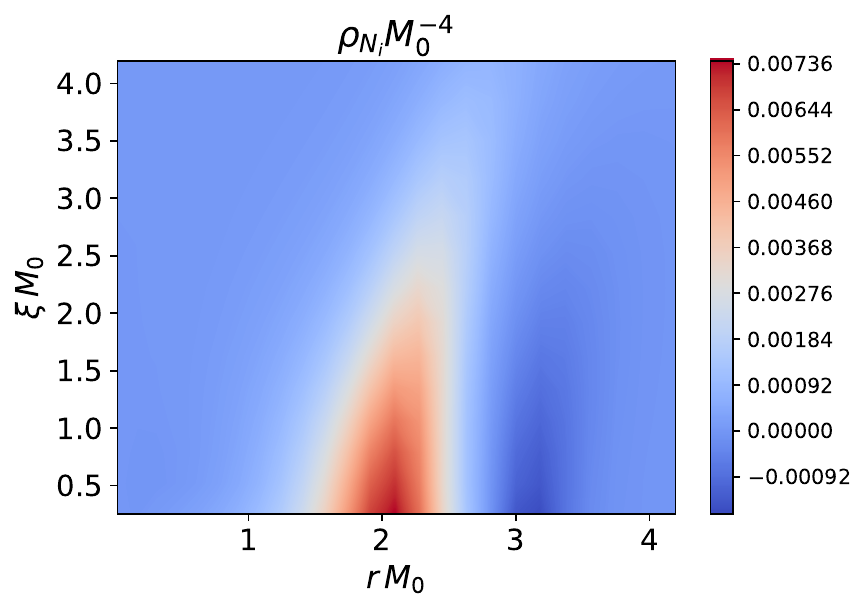}%
 \hspace{3mm}\includegraphics[width=0.49\textwidth]{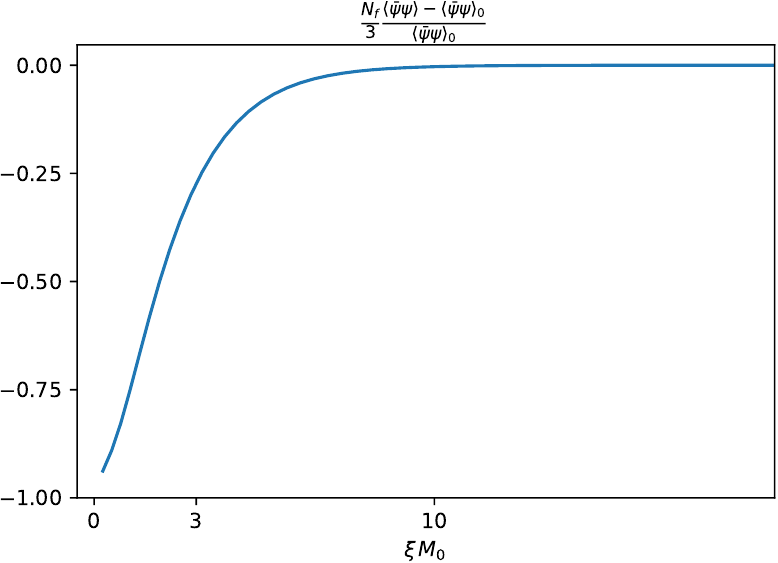}
 }
 \caption[]{Left: The instanton number density of the V-QCD baryon as a function of the holographic coordinate $r$ and the distance from the center of the baryon $\xi$. Right: The modification of the chiral condensate inside the baryon. Both figures from~\cite{Jarvinen:2022gcc}.}
 \label{fig3}
\end{figure*}

In order to solve the soliton we write an ansatz which respects the expected symmetries of the solution and choose an appropriate gauge, following the setup in hard wall models~\cite{Pomarol:2007kr,Pomarol:2008aa,Gorsky:2013dda,Gorsky:2015pra}. Also the boundary conditions need to be figured out, which is somewhat nontrivial as the baryon is a topological soliton: information on the winding of the soliton is carried by the tail configuration of the gauge fields and the phase of the tachyon, which vanish slowly far away from the center of the soliton~\cite{Jarvinen:2022mys}. We restricted to solutions only depending on the holographic coordinate and a single spatial coordinate, i.e., the distance from the baryon center. The resulting nonlinear partial differential equations could then be solved by using a relaxation method~\cite{Jarvinen:2022gcc}.

The result for the bulk instanton density $\rho_{N_i}$ is shown in Fig.~\ref{fig3} (left) as a function of the holographic coordinate $r$ and the spatial radial coordinate $\xi$. All quantities are given in units of the classical soliton mass $M_0$. As expected, the center of the soliton, where highest densities are found, is located at a finite value of the holographic coordinate, $r \approx 2/M_0$. The interpretation is that the soft wall, created by the geometry, and most importantly the tachyon field in the IR, stops the baryon from falling to the bottom of the space.

We also solved to leading order in $1/N_f$ the backreaction of the soliton solution to the tachyon~\cite{Jarvinen:2022gcc}. From this one can solve how the chiral condensate is modified due to the presence of the baryon. The result is shown in Fig.~\ref{fig3} (right) as a function of the distance from the soliton center. Chiral symmetry is partially restored inside the baryon, which is the expected behavior.

\begin{table}[h]
\centering
\begin{tabular}{|c|c|c|}
\hline
Spin & V-QCD mass & Experimental mass \\
\hline
$s = \frac{1}{2}$ & $M_N \simeq 1170 \,\text{MeV} $ & $M_N = 940 \,\text{MeV} $\\
\hline
$s = \frac{3}{2}$ & $M_\Delta \simeq 1260 \,\text{MeV}$ & $M_\Delta = 1234 \,\text{MeV}$
 \\
\hline
\end{tabular}
\caption{The masses of the nucleons and $\Delta$ baryons in the model compared to experimental data.}
\label{tab:masses}
\end{table}

We also considered slow rotation of the soliton, from which one can compute the moment of inertia of the baryon and consequently the mass of the $\Delta$ baryons~\cite{Panico:2008it,Cherman:2011ve}. This requires solving a set of linear partial equations,
The results for the masses of the nucleons and the $\Delta$ baryons are given in Table~\ref{tab:masses}.

\section{Holographic nuclear matter}\label{sec:NM}

I now move to the discussion of nuclear matter, i.e., a dense phase of QCD matter formed (mostly) out of tightly packed nucleons. To describe such a phase using holography, one should in principle consider a highly inhomogeneous soliton crystal, constructed by putting together a large number of solitons such as those discussed in Sec.~\ref{sec:soliton}. This is highly technical, and one encounters an additional difficulty: at large $N_c$ the infinitely heavy baryons indeed form a crystal, whereas for $N_c=3$ nuclear matter is expected to be a superfluid Fermi liquid~\cite{Kaplunovsky:2010eh}. While some developments studying such crystal exists in the literature~\cite{Kim:2007vd,Rho:2009ym,Kaplunovsky:2012gb,Kaplunovsky:2013iza,Kaplunovsky:2015zsa,Jarvinen:2020xjh}, due to the aforementioned issues, no full fledged three or four dimensional soliton crystal solution has been found in holographic models. It has also been proposed that the instantons break into a lattice of half-instantons above a critical density (see~\cite{Goldhaber:1987pb,Kugler:1988mu,Park:2002ie,Lee:2003aq,Rho:2009ym,Lee:2015qsa,Paeng:2015noa}). Possibly related transitions have been studied in the WSS model in~\cite{Elliot-Ripley:2016uwb,Kovensky:2020xif,CruzRojas:2023ugm}. 

In the rest of the article we will bypass the issues listed above by using a simple approximation scheme, where nuclear matter is modeled through a homogeneous field in the bulk~\cite{Rozali:2007rx,Li:2015uea,Elliot-Ripley:2016uwb,Kovensky:2021ddl,Bartolini:2022rkl,CruzRojas:2023ugm,Kovensky:2023mye}. This scheme also assumes $N_f=2$, but it can be easily generalized to describe a two-flavor subsector embedded into a higher dimensional flavor space. In the case of V-QCD, the homogeneous Ansatz is written as
\be
 A_L^i = -A_R^i = h(r) \sigma^i
\ee
where $\sigma^i$ are the Pauli matrices. It therefore involves the spatial components of the non-Abelian gauge fields with fixed parity. This approach however has an additional issue: for smooth $h(r)$, the baryon number of the solution is always zero. This can be fixed by introducing a discontinuity in the bulk~\cite{Rozali:2007rx}. Such a discontinuity can also be motivated by smearing solitons (in singular Landau gauge), i.e., integrating over their spatial locations~\cite{Jarvinen:2021jbd}. After this, thermodynamics of the nuclear matter phase can be analyzed by following the standard holographic dictionary. For V-QCD, the phase structure is shown in Fig.~\ref{fig2} where the nuclear matter phase (at low temperature and intermediate density) is shown in blue color~\cite{Ishii:2019gta}.

Interestingly, the EOS in the nuclear matter in this approach is stiff: the speed of sound is relatively high, satisfying $c_s^2>1/3$ in the high density end of the nuclear matter phase. This is important as it makes it easier to construct models that pass the astrophysics bounds from neutron star observations that I discussed above.

\begin{figure} 
 \centerline{\includegraphics[width=0.45\textwidth]{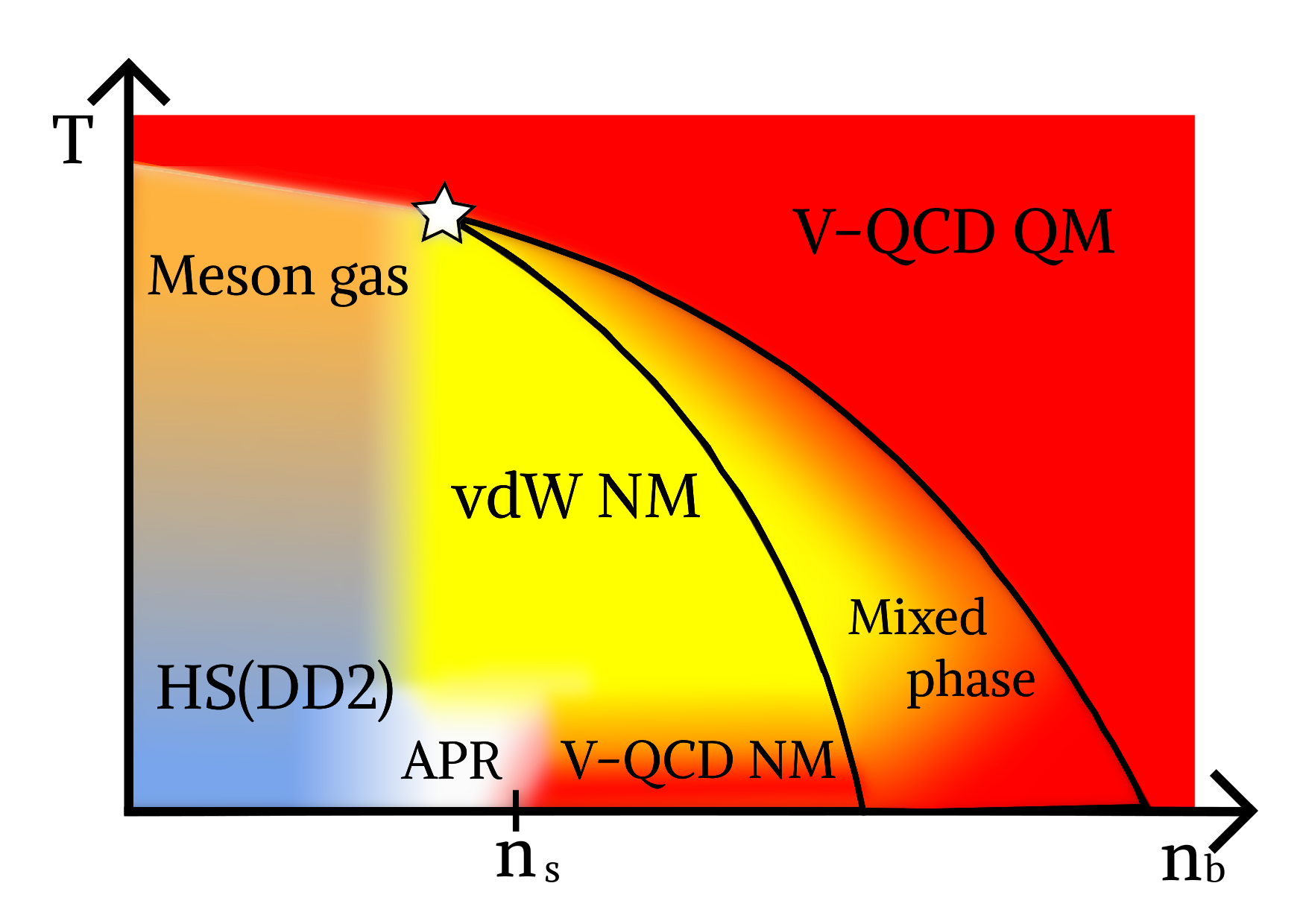}}
 \caption[]{Sketch of the regions of the phase diagram at finite temperature and density which are described by using different methods in the ``hybrid'' approach. From~\protect\cite{Demircik:2021zll}.}
 \label{fig4}
\end{figure}

\section{``Hybrid'' Equations of State} 

The V-QCD EOS, which follows from the model as constructed above, has some obvious shortcomings:
\begin{itemize}
 \item The homogeneous nuclear matter approach is only expected to work at high densities. At low densities, treatment based on individual nucleons should be more appropriate.
 \item The temperature dependence of thermodynamics in the thermal gas and nuclear matter phases is trivial: all thermodynamic quantities only depend on the chemical potential.
\end{itemize}

In order to cure the former issue in holography, one should consider many instanton solutions, but as I pointed out above, this is highly challenging. Another option would be to simply use the homogeneous phase as such, and at low densities, consider a mixed phase of vacuum and homogeneous holographic nuclear matter~\cite{Schmitt:2020tac,Kovensky:2021kzl}, which actually leads to rather realistic low density EOS. We have however chosen yet another approach: at low densities, it is not necessary to use holography at all, because the EOS is know to a good precision from ``traditional'' nuclear theory methods, such as chiral effective theory. Therefore we will simply use effective theory at low density and V-QCD at high density, which gives rise to a ``hybrid'' EOS~\cite{Ecker:2019xrw,Jokela:2020piw,Demircik:2021zll}.  

The latter issue arises because of the implicit use of large $N_c$: the temperature dependence would arise from string loop contributions that are suppressed by power of $1/N_c$. Including such corrections on the bulk side would also be highly challenging. At low densities we can cure the issue by using similar idea as for the density dependence: simply use effective theory. At higher densities in the nuclear matter phase the issue is however more severe as there are no known reliable methods to estimate the temperature dependence. In the absence of such methods, we will use a simple estimate based on a van-der-Waals EOS~\cite{Demircik:2021zll}. The goal of this construction is to produce a model that passes known theoretical and observational constraints, and is general enough to be used in state-of-the-art neutron start binary merger simulations.

As a result, the most developed hybrid V-QCD finite temperature EOS~\cite{Demircik:2021zll} has the structure depicted in Fig.~\ref{fig4}. Apart from V-QCD (red regions) the building blocks include the following:
\begin{itemize}
 \item At low densities in the nuclear matter phase, we use the Hempel-Schaffner-Bielich (HS) model~\cite{Hempel:2009mc} with DD2 interactions~\cite{Typel:2009sy}. This model is based on a statistical model below the nuclear saturation density, and mean field theory above it. We also add the pressure of free gas of mesons from the particle data group listings~\cite{ParticleDataGroup:2022pth}, including all states with masses up to 1~GeV.
 \item For the temperature dependence in the dense nuclear matter phase we use a van der Waals model, i.e., a gas of protons, neutrons and electrons with excluded volume correction for the nucleons and mean field potential between them (see, e.g.,~\cite{Rischke:1991ke,Vovchenko:2016rkn,Vovchenko:2017zpj}). The potential is tuned in such a way that the model exactly reproduces the V-QCD nuclear matter EOS at zero temperature. Therefore it provides an extrapolation of the V-QCD results to higher temperature in the nuclear matter phase.
 \item Near the saturation density, we use the Akmal-Pandharipande-Ravenhall model~\cite{Akmal:1998cf} at zero temperature. This is because both the V-QCD and HS(DD2) EOSs are rather stiff, and combining them directly would lead to families of models that would have trouble with passing the tidal deformability constraint from GW170817. 
\end{itemize}
A state-of-the-art EOS also needs to be able to describe matter out of $\beta$-equilibrium as such nonequilibrium conditions are known to develop in mergers. In practice, this means that dependence on electron fraction $Y_e$ is needed. For the strongly coupled sector this maps due to charge neutrality to the proton fraction, i.e., the number of protons over the baryon number. As it turns out, the dependence on $Y_e$ predicted by the van der Waals model is not realistic enough: the symmetry energy~\cite{Baldo:2016jhp}  is smaller than seen in experiments. Therefore we use the HS(DD2) model for the $Y_e$ dependence in the nuclear matter phase. 

We choose three representative EOSs (soft, intermediate, stiff) which reflect the parameter dependence of the V-QCD model that is left unfixed by the comparison to lattice data. These EOSs are publicly available in the CompOSE database~\cite{Typel:2013rza,CompOSECoreTeam:2022ddl}. They are in agreement with all known theoretical and observational constraints. In particular, they agree with the available measurements of masses and radii of neutron star within their margins of error, including the recent results from the NICER experiment for the radius of the massive pulsar J0740+6620~\cite{Miller:2021qha,Riley:2021pdl}. However, the comparison works a bit better for the intermediate and stiff models than the soft model~\cite{Jokela:2020piw,Demircik:2020jkc,Jokela:2021vwy,Demircik:2021zll}.

An important feature of these EOSs is a strong first order phase transition from cold nuclear matter to quark matter, which is almost completely (apart from the weak temperature dependence in the nuclear matter phase) described by the V-QCD model. Because of the strong transition, the model predicts that isolated neutron stars do not have quark matter cores~\cite{Jokela:2018ers,Ecker:2019xrw,Jokela:2020piw}. At higher temperatures, however, the transition becomes weaker. After the inclusion of the pressure of meson gas, a clear critical point can be identified. At the critical point we find that
\begin{align}
 &110~\mathrm{MeV} \lesssim T_c \lesssim 130~\mathrm{MeV}  &\\ &480~\mathrm{MeV} \lesssim \mu_{bc} \lesssim 580~\mathrm{MeV} &
\end{align}
where the uncertainties arise from varying the model between the three different versions (soft, intermediate, and stiff). The location of the critical point is close to what is obtained in simpler holographic models that are also fitted to lattice data~\cite{DeWolfe:2010he,Knaute:2017opk,Critelli:2017oub,Cai:2022omk,Li:2023mpv}.

\section{Application to neutron star mergers}

Gravitational wave signals from the collision of two neutron star were recently observed by the LIGO/Virgo collaboration. By far the cleanest event so far is the first event, GW170817, for which the electromagnetic counterpart was also observed at essentially all possible wavelengths~\cite{LIGOScientific:2017zic,GBM:2017lvd}. 

The merger events are expected to fall into three main categories (see, e.g.,~\cite{Baiotti:2016qnr}). 
\begin{enumerate}
 \item If the neutron stars have high masses, the system collapses into a black hole promptly after the merger. In this case the gravitational wave signal contains an inspiral phase and a rapid ringdown. 
 \item At intermediate total mass, a short-lived hypermassive neutron star is formed, with the mass well above the maximum for a nonrotating star. Initially the star is differentially rotating, but as the differential rotation declines, the remnant collapses into a black hole within one second of the merger or so. In this case, the gravitational wave signal has a potentially interesting additional phase after the merger, driven by the oscillations of the hypermassive remnant, which has higher typical frequencies than the inspiral phase.
 \item  At lower masses, the neutron star remnant does not collapse at least at short timescales, i.e. within seconds. A collapse at longer timescales as the star cools down and rotation slows, is still possible. In this case, the gravitational wave signal contains an inspiral phase and aftermerger oscillation phase, but no signal of collapse.
\end{enumerate}

\begin{figure*} 
 \centerline{\includegraphics[width=0.9\textwidth]{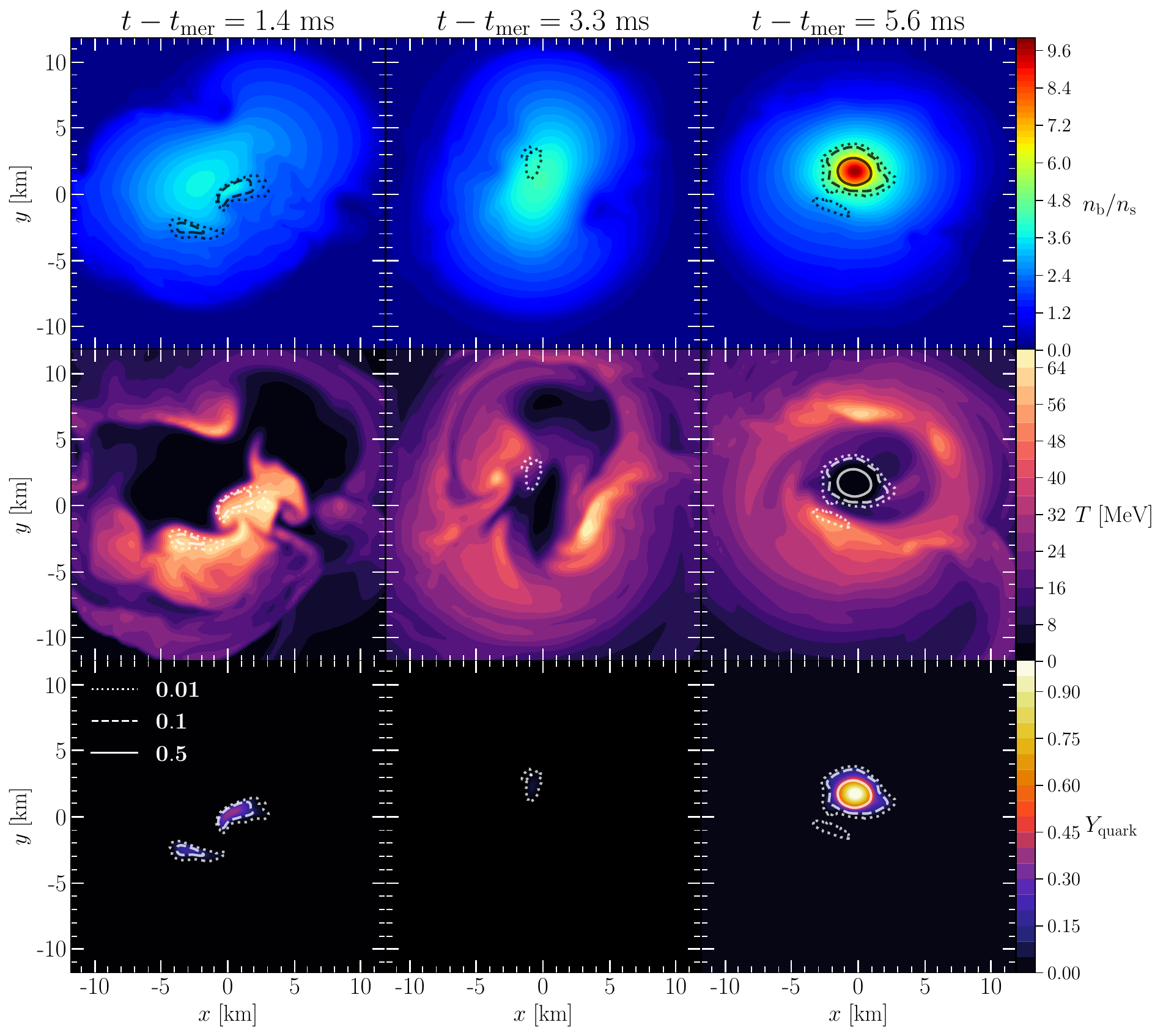}}
 \caption[]{Snapshots of the state of the hypermassive neutron star formed in a binary neutron star collision in the equatorial plane at three different stages after the merger in different columns. The top, middle, and bottom rows show the density, temperature, and quark fraction, respectively. The additional contours show the regions where quark matter is formed.   From~\protect\cite{Tootle:2022pvd}.}
 \label{fig5}
\end{figure*}

We studied neutron star mergers by using the hybrid V-QCD EOSs~\cite{Demircik:2021zll} in~\cite{Tootle:2022pvd}. Simulating neutron star mergers requires solving the evolution of 3+1 dimensional general relativity coupled to hydrodynamics. This is a numerically challenging problem which requires supercomputing. We used Frankfurt University/Kadath (FUKA) spectral code for initial data~\cite{Papenfort:2021hod} and Frankfurt/Illinois (FIL) code for binary evolution~\cite{Most:2019kfe}, both implemented in the Einstein toolkit framework. Simulations were carried out as a part of the project ``BNSMIC'' 
on HAWK supercomputer at the High-Performance Computing Center Stuttgart.

As the hybrid V-QCD EOSs are one of the few examples of state-of-the-art models which include controlled predictions for the nuclear to quark matter phase transition, we focused on details of quark matter production for mergers with total mass determined by the GW170817 event (see also~\cite{Most:2018eaw,Bauswein:2018bma,Prakash:2021wpz}). Analysis of the electromagnetic signal from this event suggests that a hypermassive neutron star was formed which collapsed into a black hole about one second after the merger. I show snapshots from a simulation using the soft hybrid EOS in Fig.~\ref{fig5}. In this figure, the top, middle, and bottom rows show the density, temperature, and quark fraction in the equatorial plane, respectively. The first, second, and third columns show the state of the hypermassive neutron star at 1.4, 3.3, and 5.6 milliseconds after the merger, respectively. These snapshots were chosen so that they present three different stages of quark matter production: hot, warm, and cold quarks. The characteristics of these stages are the following:
\begin{itemize}
 \item \emph{Hot quarks} are produced in the early evolution after the merger, as the heating of the QCD matter takes it above the transition line while densities remain relatively low. Hot quarks typically appear in the hottest regions of the simulation, as is the case in the left column of Fig.~\ref{fig5}.
 \item \emph{Warm quarks} are the result of complicated oscillating dynamics and appear in regions that are neither hottest nor densest in the remnant. 
 \item \emph{Cold quarks} appear in the center of the star where the density is at its highest at late stages. The formed quark matter has essentially zero temperature, apparently due to the large latent heat required to transform from nuclear to quark matter (see the right column in Fig.~\ref{fig5}). Formation of cold quark matter eventually leads to a collapse to a black hole as there are no stable stars with quark matter codes in the models we are using as the quark matter EOS is rather soft (low speed of sound).
\end{itemize}

The phase transition is also visible in the  gravitational wave analysis. The simulations where the quark matter component is removed form the EOS by hand typically show much longer lifetimes for the remnant before collapse to the black hole, or no collapse at all. Moreover for the soft EOS we find that the lifetime of the remnant $\sim 10$~ms is much smaller than the estimate~\cite{Gill:2019bvq} that the collapse for GW170817 took place about one second after the merger. This observation disfavors the soft variant of the EOS.

\section{Conclusion}

I reviewed  the V-QCD model, focusing on topics related to baryons, nuclear matter, and applications to neutron star physics. 

I started with a discussion of the single baryon solution in the V-QCD model~\cite{Jarvinen:2022mys,Jarvinen:2022gcc}, i.e., a new soliton solution for the gauge fields in the holographic dual. This solution had various desired features that were in part missing in earlier solutions found in the hard wall and WSS models: the location of the center of the soliton was dynamically determined in a consistent model of the IR physics, and the coupling of the soliton to the chiral symmetry breaking was included. The masses of the nucleons and the $\Delta$-baryons agree well with experiments, and that the chiral condensate is partially restored inside the baryon.

I also discussed the phase diagram and the EOS. I demonstrated that the V-QCD EOS is realistic enough in order to construct, in combination with other models (in our case mostly the HS(DD2) nuclear theory model), a state-of-the-art model for neutron star EOS which is in agreement with known theoretical and observational constraints. This was possible thanks to a few main successes in the model:
\begin{itemize}
 \item A precise fit to lattice data was possible~\cite{Jokela:2018ers}, leading to a realistic extrapolation of the quark matter EOS to high densities.
 \item By using the homogeneous model for nuclear matter, it was possible to describe both nuclear and quark matter in a single framework. In particular, the properties of the phase transition could be predicted~\cite{Ishii:2019gta,Jokela:2020piw,Demircik:2021zll}.
 \item The EOS in the nuclear matter was seen to be stiff, with the values of the speed of sound easily exceeding the conformal value $c_s^2=1/3$~\cite{Ishii:2019gta}. This helped in the construction of feasible EOSs, because observational constraints (high values of some of the reliably measured neutron star masses) favor stiff EOSs~\cite{Ecker:2019xrw,Jokela:2018ers,Jokela:2020piw,Jokela:2021vwy}.
\end{itemize}

By using V-QCD in combination with the HS(DD2) model and a van der Waals model, it was possible to derive EOSs which included both realistic temperature dependence and dependence on the electron fraction~\cite{Demircik:2021zll}. After this the EOSs could be directly used in state-of-the-art neutron star simulations~\cite{Tootle:2022pvd}. As the EOSs contain both nuclear and quark matter in a single framework, it was natural to focus on the phase transition and the production of quark matter in the merger. We identified three different stages of quark matter production: initially ``hot quarks'' in the hottest regions of the remnant, thereafter ``warm quarks'' due to violent oscillations of the remnant, and eventually ``cold quarks'' in the cold and dense core of the formed hypermassive neutron star.

There are various ongoing and planned future projects which aim at improving the EOS and extending the analysis of these holographic models towards new directions. These include the following: 
\begin{itemize}
 \item \emph{Extensions of the EOS.} I am planning to improve the V-QCD EOS by adding proper implementations of quark flavors, including the effects due to the strange quark mass and improved analysis of the flavor-asymmetric configurations, i.e., symmetry energy. Some of these aspects have already been studied in the WSS and hard wall models~\cite{Kovensky:2019bih,Kovensky:2021ddl,Bartolini:2022gdf,Kovensky:2023mye}. Another direction would be to study the dependence on the magnetic field~\cite{Gursoy:2017wzz,Gursoy:2020kjd} with properly defined, flavor dependent electric current. Moreover I am planning to include a model for a paired, color superconducting phase which will modify the results at high densities.
 \item \emph{Transport.} Analysis of viscosities and conductivities in the V-QCD quark matter phase already appeared in~\cite{Hoyos:2020hmq,Hoyos:2021njg}. A very recent article studied neutrino transport in strongly coupled holographic plasma using a simple holographic model~\cite{Jarvinen:2023xrx}. Future work will extend the analysis to the bulk viscosity due to weak interactions coupled to QCD matter, and to neutrino transport in the full V-QCD model.   
 \item \emph{Domain walls.} Even if there are no stable quark matter cores, domain walls between the nuclear and quark matter phases appear in hypermassive neutron stars formed in mergers. Ongoing work will solve domain walls between the vacuum and nuclear matter phase, but this work can be extended to the walls between nuclear and quark matter. From these solutions one can extract, among other things, the surface tension of the domain wall.   
\end{itemize}

\section*{Acknowledgment}
I thank the organizers of the 6th International Conference on Holography, String Theory and Spacetime in Da Nang, for the invitation to give a talk, and the participants of the conference for interesting discussions.
This research has been supported by an appointment to the JRG Program at the APCTP through the Science and Technology Promotion Fund and Lottery Fund of the Korean Government and by the Korean Local Governments -- Gyeong\-sang\-buk-do Province and Pohang City -- and by the National Research Foundation of Korea (NRF) funded by the Korean government (MSIT) (grant number 2021R1A2C1010834).

\bibliographystyle{JHEP}
\bibliography{mj-biblio.bib}

\end{document}